\documentstyle[prl,aps,epsf,twocolumn]{revtex}

\begin{document}

\title{A New Strategy of Quantum-State Estimation\\ for Achieving the Cram\'{e}r-Rao Bound}
\author{Koji Usami$^{1,2}$~\cite{email}, Yoshihiro Nambu$^{2,3}$, Yoshiyuki Tsuda$^{4}$, \\
Keiji Matsumoto$^{4}$, and Kazuo Nakamura$^{1,2,3}$}
\address{$^1$Depertment of Material Science and Engineering, Tokyo Institute of Technology, 4259 Nagatsuta-cho, Midori-ku, Yokohama, Kanagawa, 226-0026,Japan}
\address{$^2$CREST, JST, 3-13-11 Shibuya, Shibuya-ku, Tokyo, 150-0002, Japan}
\address{$^3$Fundamental Research Laboratories, NEC, 34 Miyukigaoka, Tsukuba, Ibaraki, 305-8501, Japan}
\address{$^4$ERATO, JST, 5-28-3 Hongo, Bunkyo-ku, Tokyo, 113-0033, Japan}
\date{\today}
\maketitle

\begin{abstract}
We experimentally analyzed the statistical errors in quantum-state estimation and examined whether their lower bound, which is derived from the {\it Cram\'{e}r-Rao inequality}, can be truly attained or not. In the experiments, polarization states of bi-photons produced via spontaneous parametric down-conversion were estimated employing tomographic measurements. Using a new estimation strategy based on {\it Akaike's information criterion}, we demonstrated that the errors actually approach the lower bound, while they fail to approach it using the conventional estimation strategy.
\end{abstract}

\vspace{0.25cm}


\vspace{0.25cm}


One of the central features of quantum mechanics is that it does not allow obtaining complete information about an individual quantum system without errors~\cite{MP1995}. The {\it Holevo bound} and the {\it no-cloning theorem} are prominent manifestations of the restrictions on acquiring information from a quantum system~\cite{NCtext2000}. Although it is possible to estimate all aspects of a quantum state by performing a series of distinct measurements on identically prepared particles, that is, by tomographic measurements~\cite{Leonhardt1997,WJEK1999,JKMW2001}, there is still a lower bound on the statistical errors. Therefore, it is important to establish an estimation strategy that can attain the lower bound to improve the accuracy and sensitivity of the precision measurements that exploit quantum states~\cite{HB1993,SM1995,BK1997} and to develop quantum information technology that requires us to faithfully prepare quantum states~\cite{NCtext2000}.

In this article, we report our experimental analysis of the statistical errors in quantum-state estimation. Various polarization states of bi-photons produced via spontaneous parametric down-conversion were estimated employing tomographic measurements~\cite{JKMW2001}. Although estimating such a two-qubit system has already been investigated in detail by James, {\it et al.}~\cite{JKMW2001}, achieving the lower bound on statistical errors, which is derived from {\it Cram\'{e}r-Rao inequality}, is highlighted in this article. We experimentally examined whether the lower bound was truly achieved, and found that there are situations in which the use of redundant parameters to estimate the quantum state leads to ambiguities in the estimates and prevents the errors from approaching their lower values. We then developed a new strategy for estimating the quantum state based on {\it Akaike's information criterion (AIC)}~\cite{Akaike1974}, which provides a rigorous way of selecting the most suitable number of parameters characterizing a quantum state. Using the new estimation strategy, we found that the errors in the quantum-state estimation actually approach the lower bound. 

\begin{figure}
\epsfxsize=8cm
\epsfbox{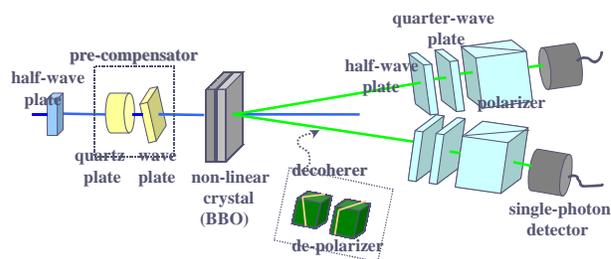}
\caption{Experimental setup for producing various polarization states of bi-photons and measuring them.}
\label{fig:setup}
\end{figure}

A rough sketch of the experimental setup is shown in Fig.~\ref{fig:setup}. We generated bi-photons at a wavelength of 532~nm via type-I spontaneous parametric down-conversion (SPDC) with {\it two} beta-barium-borate (BBO) crystals attached together~\cite{KWWAE1999}. The method used to produce various polarization states of bi-photons was almost the same as that used in Refs.~\cite{WJEK1999,WJMK2001}; the major difference was that we used an ultra-short pulse laser (the third harmonics of a mode-locked Ti:Sapphire laser) as a pump beam for the SPDC~\cite{NUTMN2002}. We produced three quantum states, {\it almost pure separable state (APSS)}, {\it highly entangled state (HES)}, and {\it very noisy mixed state (VNMS)}, by adjusting the pump-beam polarization with a half-wave plate~\cite{WJEK1999}, by modifying the relative time delay between the horizontal and vertical components of the pump beam with a {\it pre-compensator}~\cite{NUTMN2002}, and by inserting {\it decoherers}, i.e., two de-polarizers, into either paths of the down-converted photons~\cite{WJMK2001}. The coincident (within 6~ns) detection events, $n_{\nu}$, on both single-photon detectors (HAMAMATSU H7421-40) at a certain polarizer's setting (i.e., projector) $|m_{\nu}\rangle \langle m_{\nu}|$ (which was determined by the half-wave plate, quarter-wave plate, and polarizer on each path of the converted photons) were counted by using a time interval analyzer (YOKOGAWA TA-520) during data acquisition time $t$. To eliminate ambient photons, interference filters (FWHM: 8~nm) were used. The typical coincidence counting rate was about 500~Hz. 

The polarization states of the bi-photons (two-qubit states) can be characterized by a $4 \times 4$ density matrix with 15 independent real parameters, $\rho_{\Theta}=T_{\Theta}T_{\Theta}^{\dagger}/{\rm Tr}[T_{\Theta}T_{\Theta}^{\dagger}]$, where $T_{\Theta}$ is the complex lower triangular matrix with 16 real parameters $\{\theta^{\mu}\}_{\mu=1}^{16} \equiv \{\Theta\}$~\cite{JKMW2001,BDPS1999}. To determine these parameters, coincidence counting measurements at 16 particular polarizers settings, $\{|m_{\nu}\rangle \langle m_{\nu}|\}_{\nu=1}^{16}$~\cite{note_tomography}, were used as the {\it tomographic measurements}~\cite{JKMW2001}. From these coincidence counts, $\{n_{\nu}\}_{\nu=1}^{16} \equiv \{N\}$, parameters $\{\Theta\}$ and thus quantum state $\rho_{\Theta}$ were estimated using maximum likelihood estimation (MLE)~\cite{JKMW2001,BDPS1999,Hradil1997,RHJ2001}.

We assessed the errors in the estimation as follows. If the estimation procedure is repeated $r$ times under identical conditions, we have $r$ slightly different quantum states, $\{\rho_{\hat{\Theta_{i}}}\}_{i=1}^{r}$, due to statistical errors. Let the {\it true} quantum state be $\rho_{\Theta_{0}}$, where $\{\Theta_{0}\}$ is the {\it true} values of the parameters. Then we calculate the average {\it Bures distance}~\cite{BC1994,Holevo2001} between the {\it true} state and the estimated states, i.e., $2(1-\bar{F}(\rho_{\Theta_{0}},\rho_{\hat{\Theta}})) \equiv 2(1-\frac{1}{r}\sum_{i=1}^{r}F(\rho_{\Theta_{0}},\rho_{\hat{\Theta_{i}}}))$, where $F(\rho_{\Theta_{0}},\rho_{\hat{\Theta_{i}}}) \ (\equiv {\rm Tr}[\sqrt{\sqrt{\rho_{\Theta_{0}}}\rho_{\hat{\Theta_{i}}}\sqrt{\rho_{\Theta_{0}}}}\,])$ is the fidelity~\cite{NCtext2000}. In the experiments, we repeated the estimation procedure nine times (i.e., r=9). The estimated quantum state using all the data for the nine trials, i.e., $\{\sum_{i=1}^{9}n_{\nu\,(i)}\}_{\nu=1}^{16} \equiv \{N_{0}\}$, is assumed to be the {\it true} state, $\rho_{\Theta_{0}}$~\cite{UNTMN}. Keeping the experimental conditions identical, the average Bures distances were measured for five data acquisition times ($t=$ 0.2~s, 0.5~s, 1~s, 2~s, 5~s). 

\begin{figure}
\epsfxsize=8.2cm
\epsfbox{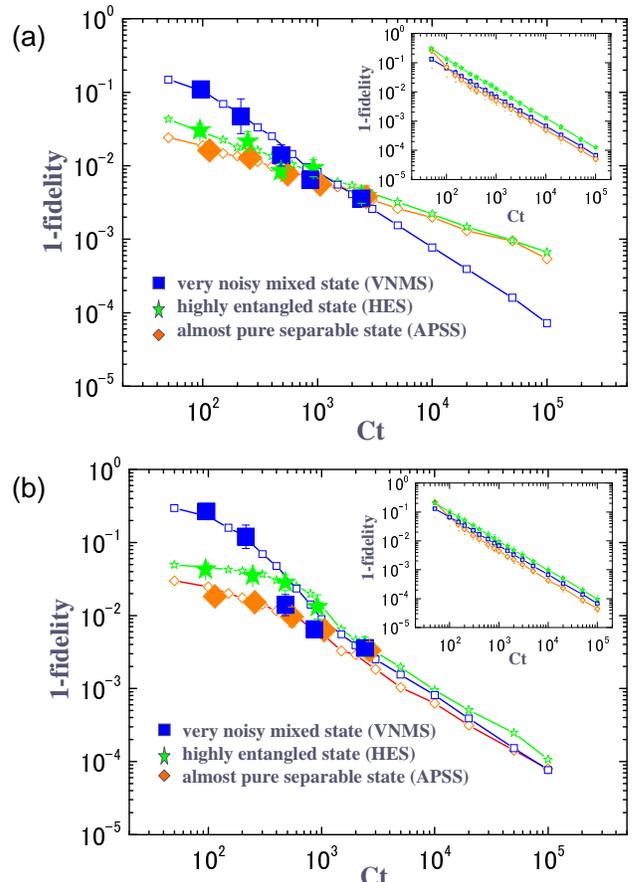}
\caption{Average Bures distances (divided by 2) between {\it true} states and states estimated by MLE~(a) and by MAICE~(b), as a function of data acquisition time $t$ multiplied by nuisance parameter $C \ (\approx 500$). The solid points (vertical error bars correspond to one standard deviation) represent experimental results and the empty points represent numerical results. The insets show the asymptotic lower bounds explained in the text.}
\label{fig:bures}
\end{figure}

Experimentally measured average Bures distances (divided by 2) between the estimated states and their {\it true} states (solid points) and those obtained by Monte Carlo simulation (empty points) are shown in Fig.~\ref{fig:bures}~(a) as a function of the data acquisition time multiplied by {\it nuisance parameter} $C$, which corresponds to the coincidence counting rate without any polarizers (i.e., $|m_{\nu}\rangle \langle m_{\nu}|=I$)~\cite{note_threestate}. The simulations were carried out by artificially producing the 16 coincidence count data $\{N\}$ according to the aforementioned {\it true} states. We estimated the states from the simulated data and repeated this procedure 200 times (i.e., r=200). Therefore, the values presented in Fig.~\ref{fig:bures} (empty points) correspond to $1-\frac{1}{200}\sum_{i=1}^{200}F(\rho_{\Theta_{0}},\rho_{\hat{\Theta_{i}}})$. The results of the experiments and numerical simulations are in good agreement, even though the errors in experimentally estimating quantum state may stem not only from a statistical origin, but also from systematic errors in the experiment itself~\cite{note_systematic}.

Next we examined whether the theoretically expected lower bound~\cite{note_lowerbound} on the Bures distance was attained or not in each experiment. The lower bounds are shown in the inset of Fig.~\ref{fig:bures}~(a). They can be derived as follows~\cite{UNTMN}. Suppose that the estimated state is in the neighborhood of the {\it true} state, so that the Bures distance between them can be written as~\cite{BC1994,Holevo2001,Uhlmann1993,MatsumotoPhD,Hayashi2002}
\begin{equation}
2(1-\bar{F}(\rho_{\Theta_{0}},\rho_{\hat{\Theta}})) \approx \frac{1}{4}\sum_{i=1}^{16}\sum_{j=1}^{16}J_{ij}^{S}(\Theta_{0})\,V^{ij}, \label{eq:Bures}
\end{equation}
where $[J_{ij}^{S}(\Theta)] \equiv {\rm J}^{S}(\Theta)$ is the so-called {\it symmetric logarithmic derivative (SLD) Fisher information matrix}~\cite{note_SLDFisher} and $[V^{ij}]=[(\hat{\theta}^{i}(N)-\theta_{0}^{i})(\hat{\theta}^{j}(N)-\theta_{0}^{j})] \equiv {\rm V}$ is the covariance matrix. This means that the Bures distance can be locally viewed as a distance on a Riemannian manifold equipped with a metric structure defined by the SLD Fisher information matrix~\cite{BC1994,Holevo2001,Uhlmann1993,MatsumotoPhD,Hayashi2002}. 

The Cram\'{e}r-Rao inequality~\cite{CTtext1991,ANtext2000} provides an asymptotic lower bound (CR bound) on covariance matrix ${\rm V}$ in Eq.~(\ref{eq:Bures}): ${\rm V} \geq {\rm J}^{-1}(\Theta)|_{\Theta=\Theta_{0}}$, where ${\rm J}(\Theta)$ is {\it Fisher's information matrix}~\cite{note_Fisher}. The CR bound tells us that no estimation strategy exceeds this bound in the asymptotic region. With a little calculation, we have ${\rm J}(\Theta) \propto C t$~\cite{note_Fisher}. Therefore, we can define $\tilde{{\rm J}}(\Theta)$ as ${\rm J}(\Theta) \equiv C t \tilde{{\rm J}}(\Theta)$. Consequently, from Eq.~(\ref{eq:Bures}), the asymptotic lower bound on the Bures distance between the {\it true} state and the estimated state is given by
\begin{equation}
2(1-\bar{F}(\rho _{\Theta _{0}},\rho _{\hat{\Theta}})) \geq \frac{1}{4 C t}{\rm Tr}[{\rm J}^{S}(\Theta_{0})\,(\tilde{{\rm J}}^{-1}(\Theta)|_{\Theta=\Theta_{0}})].  \label{eq:limitTM}
\end{equation}
We can see that the lower bound decreases as inversely proportional to the data acquisition time as shown in the inset of Fig.~\ref{fig:bures}~(a). 

It is known that the CR bound can be asymptotically achieved by using MLE~\cite{ANtext2000,Braunstein1992}; however, the experimental data in Fig.~\ref{fig:bures}~(a), especially those for HES and APSS, show that the average Bures distances did not decrease as inversely proportional to $t$, so the CR bound was not achieved in these experiments. This means we need an estimation strategy other than MLE.

To clarify the necessity of a new strategy, we briefly review the estimation strategy used in the above experiments. We define the {\it true} probability density function (PDF) obtaining the 16 coincidence counts $\{N\}$ (with $t=1~s$ for each) as $P(N|\Theta_{0})|_{t=1} \equiv P_{0}(N)|_{t=1}$ and its parametric model as $P(N|\Theta)|_{t=1}$. The relative entropy~\cite{CTtext1991,ANtext2000} between them is
\begin{eqnarray}
&\!& D(P_{0}(N)\,\Vert \,P(N|\Theta))|_{t=1}  \nonumber \\
& \quad & =\sum_{n_{1}=0}^{\infty}\!\cdots\!\sum_{n_{16}=0}^{\infty } (P_{0}(N)\ln [P_{0}(N)]  \nonumber \\
& \quad & \quad \quad \quad \quad \quad \quad -P_{0}(N)\ln [P(N|\Theta)])|_{t=1}, \label{eq:relativeE}
\end{eqnarray}
and it takes a positive value, unless $P_{0}(N)=P(N|\Theta)$ for all $\{N\}$ (in this case, $D(P_{0}(N)\,\Vert \,P(N|\Theta))|_{t=1}=0$). Therefore, by minimizing the relative entropy in Eq.~(\ref{eq:relativeE}) with respect to $\{\Theta\}$, we can find the point $\{\hat{\Theta}(N)\}$ closest to the {\it true} point $\{\Theta_{0}\}$ in the 16-dimensional parameter space. Because the approximation 
\begin{equation}
\sum_{n_{1}=0}^{\infty}\!\cdots\!\sum_{n_{16}=0}^{\infty}\!P_{0}(N)\ln[P(N|\Theta)])|_{t=1} \approx \frac{1}{t}\ln [P(N|\Theta)]  \label{eq:Mloglikelihood}
\end{equation}
is valid as the data acquisition time $t$ is increased infinitely, this minimization is accomplished by maximizing the logarithm of the likelihood function (log-LF) $\ln[P(N|\Theta)]$. Namely, by finding parameters $\{\hat{\Theta}(N)\}$ that maximize log-LF, the quantum state can be estimated (maximum likelihood estimation or MLE~\cite{JKMW2001,BDPS1999,Hradil1997,RHJ2001})~\cite{note_findminimum}. MLE has been used for estimating the quantum states~\cite{WJEK1999,JKMW2001,NUTMN2002}. However, the ill-behavior of MLE in estimating the nearly degenerate states, such as HES and APSS, is clearly shown in Fig.~\ref{fig:bures}~(a). 

The ill-behavior can be explained as follows. Quantum state $\rho_{\Theta}$ has been implicitly assumed to be non-degenerate (i.e., in the interior of the space of the density matrix). Therefore, if the {\it true} state is a (nearly) degenerate state, such as HES or APSS, (i.e., in (the neighborhood of) the boundary of the space of the density matrix, where one or more eigenvalues vanish), the density matrix can be completely characterized using less than 15 parameters. Subsequently, the surplus parameters give rise to several local maximums of the log-LF in the 16-demensional parameter space and prevent the errors in the estimation from approaching their lower values. 

Thus, for attaining the lower bound, finding the appropriate number of parameters seems to be requisite. The {\it Akaike's information criterion (AIC)}~\cite{Akaike1974} provides a rigorous method for eliminating redundant parameters. It is defined by 
\begin{equation}
AIC^{(k)}(\Theta)=-2\times \ln[P^{(k)}(N|\Theta)]+2\times k,  \label{eq:AIC}
\end{equation}
where $k$ is the number of independent parameters and $\ln[P^{(k)}(N|\Theta)]$ corresponds to log-LF for quantum state $\rho^{(k)}_{\Theta}$. Here $\rho^{(k)}_{\Theta}$ is defined as the quantum state that is contracted within a k-dimensional parameter space. To estimate the two-qubit system, we use $\rho_{\Theta}^{(16)}$, $\rho_{\Theta}^{(15)}$, $\rho_{\Theta}^{(12)}$, and $\rho_{\Theta}^{(7)}$, which represent the rank-4, 3, 2, and 1 density matrices, respectively~\cite{UNTMN,note_parameters}. Among these four states, the one that attains the smallest AIC can be regarded as the most appropriate one for the following reason. Approximation~(\ref{eq:Mloglikelihood}), used for explaining MLE, needs a correction depending on the number of parameters~\cite{Akaike1974}; that is,
\begin{eqnarray}
& \! & \sum_{n_{1}=0}^{\infty}\!\cdots\!\sum_{n_{16}=0}^{\infty}\!P_{0}(N)\ln[P^{(k)}(N|\hat{\Theta}(N))]|_{t=1} \nonumber \\
& \quad & \quad \approx \frac{1}{t}\ln [P^{(k)}(N|\hat{\Theta}(N))] - \frac{k}{t}.  \label{eq:difference}
\end{eqnarray}
Taking this correction into account, the relative entropy, Eq.~(\ref{eq:relativeE}), can be minimized by minimizing the value, $\frac{1}{2t} AIC^{(k)}(\Theta)$. Therefore, if the quantum state that minimizes the AIC is chosen from several alternatives, this state is the one closest to the {\it true} one from the viewpoint of relative entropy ({\it minimum AIC estimate} or {\it MAICE}~\cite{Akaike1974}).

Figure~\ref{fig:bures}~(b) shows the resulting average Bures distances between the {\it true} states and the states estimated using MAICE. The Bures distances nearly reach the lower bounds given by Eq.~(\ref{eq:limitTM}) in the asymptotic regime (i.e., in the regime where $Ct \ge 10^{3}$), even when estimating degenerate states such as APSS and HES (for which the AIC mostly gave the rank-2 state as the most suitable one). Therefore, by using MAICE and thus eliminating the redundant parameters, we demonstrated that the CR bound on the statistical errors in quantum-state estimation is achievable in the asymptotic regime~\cite{note_systematic}. This means that from give experimental data, we can obtain more detailed information about quantum system by using MAICE than by any other known estimation strategy.

Finally, we note that the results shown in Fig.~~\ref{fig:bures}~(b) reveal that in the asymptotic region, the average Bures distance (i.e., the error) for the entangled state, i.e., HES, is the largest of the three states, in spite of the small entropy~\cite{note_threestate}. This may stem from the nature of tomographic measurements using only separable projectors for the estimation. Further, we numerically inspected the tomographic measurements using {\it inseparable} projectors and found that it improved the accuracy not only in estimating HES, but also in estimating VNMS~\cite{UNTMN}. The latter improvement corresponds to {\it non-locality without entanglement}~\cite{MP1995,GM2000}. The new strategy, MAICE, should make it {\it experimentally} possible to analyze such peculiar features of quantum mechanics.

In conclusion, we have demonstrated that by using a new estimation strategy based on the AIC, that eliminates superfluous parameters, the lower bound on statistical errors in estimating the quantum states can be achieved. This strategy is versatile and can be applied to other classes of multi-parameter estimation problems for quantum states or even quantum channels.

We are grateful to Tohya~Hiroshima, Satoshi~Ishizaka, Bao-Sen~Shi,
Akihisa~Tomita, Masahito~Hayashi, Masahide~Sasaki, and Osamu~Hirota for their valuable advice and encouragement. We also thank Shunsuke~Kono and
Kenji~Kazui for their technical support.



\begin{thebibliography}{99}


\bibitem[*]{email}
E-mail address: usami@frl.cl.nec.co.jp

\bibitem{MP1995}  
S.~Massar and S.~Popescu, Phys.~Rev.~Lett.~{\bf 74}, 1259 (1995).

\bibitem{NCtext2000}  
M.~A.~Nielsen and I.~L.~Chuang, {\it Quantum Computation and Quantum Information} (Cambridge University Press, Cambridge,
2000).

\bibitem{Leonhardt1997}  
U.~Leonhardt, {\it Measuring the Quantum State of Light} (Cambridge University Press, Cambridge, 1997).

\bibitem{WJEK1999}  
A.~G.~White, {\it et al.}, Phys.~Rev.~Lett.~{\bf 83} 3103 (1999).

\bibitem{JKMW2001}  
D.~F.~V.~James, {\it et al.}, Phys.~Rev.~A~{\bf 64} 052312 (2001).

\bibitem{HB1993}  
M.~J.~Holland and K.~Burnett, Phys.~Rev.~Lett.~{\bf 71} 1355 (1993).

\bibitem{SM1995}  
B.~C.~Sanders and G.~J.~Milburn, Phys.~Rev.~Lett.~{\bf 75} 2944 (1995).

\bibitem{BK1997}  
P.~Bouyer and M.~K.~Kasevich, Phys.~Rev.~A~{\bf 56} R1083 (1997).

\bibitem{Akaike1974}  
H.~Akaike, IEEE trans. Automat.Contr {\bf 19}
716 (1974).

\bibitem{KWWAE1999}  
P.~G.~Kwiat, {\it et al.}, Phys.~Rev.~A~{\bf 60} R773 (1999).

\bibitem{WJMK2001}  
A.~G.~White, {\it et al.}, Phys.~Rev.~A~{\bf 65} 012301 (2001).

\bibitem{NUTMN2002}  
Y.~Nambu, {\it et al.}, e-print quant-ph/0203115.

\bibitem{BDPS1999} 
K.~Banaszek, {\it et al.}, Phys.~Rev.~A~{\bf 61} 010304(R) (1999).

\bibitem{note_tomography}
Our set of measurements was the same as that in Ref.~\cite{NUTMN2002}.

\bibitem{Hradil1997}  
Z.~Hradil, Phys.~Rev.~A~{\bf 55} R1561 (1997).

\bibitem{RHJ2001}  
J.~\u{R}eh\'a\u{c}ek, Z.~Hradil, and M.~Je\u{z}ek,
Phys.~Rev.~A~{\bf 63} 040303(R) (2001).

\bibitem{UNTMN}
K.~Usami, {\it et al.}, in preparation.

\bibitem{note_threestate}
Entropy (von Neumann entropy) $S$ and entanglement (entanglement of formation~\cite{NCtext2000}) $E$ of the {\it true} state for VNMS, HES, and APSS, which were inferred from all nine trials, were $\{S=1.995,\, E=0\}$, $\{S=0.456,\, E=0.778\}$, and $\{S=0.212,\, E=0.032\}$, respectively.

\bibitem{note_systematic}
The systematic errors become dominant when $C t\approx 10,000$ due to the uncertainty of the polarization setting~\cite{JKMW2001} and the finite extinction ratio of the polarizers (about 30~dB). Therefore, careful alignment of the experimental setup is needed to experimentally achieve the lower bound when the data acquisition time becomes longer.

\bibitem{note_lowerbound}
This lower bound is not the {\it ultimate} lower bound, which may be, in general, achieved by using {\it generalized} quantum measurements (including all measurements permitted by the rules of quantum mechanics), but one that attained by using the tomographic measurements.

\bibitem{BC1994}  
S.~L.~Braunstein and C.~M.~Caves, Phys.~Rev.~Lett.~{\bf 72} 3439 (1994).

\bibitem{Holevo2001}  
A.~S.~Holevo, {\it Statistical Structure of Quantum Theory} (Springer, Heidelberg, 2001).

\bibitem{Uhlmann1993}  
A.~Uhlmann, Rep.~Math.~Phys.~{\bf 33}, 253
(1993).

\bibitem{MatsumotoPhD}  
K.~Matsumoto, {\it A geometrical approach to
quantum estimation theory}, Ph.D.~Thesis, The University of Tokyo (1997).

\bibitem{Hayashi2002}  
M.~Hayashi, e-print quant-ph/0202003.

\bibitem{CTtext1991}  
T.~Cover and J.~Thomas, {\it Elements of Information Theory} (Wiley, New York, 1991).

\bibitem{ANtext2000}  
S.~Amari and H.~Nagaoka, {\it Methods of Information Geometry} (AMS, Providence, 2000).

\bibitem{note_SLDFisher}
The SLD Fisher information matrix, ${\rm J}^{S}(\Theta )$, is given as follows. First, we define a Hermitian operator, ${\rm L}_{i}^{S}(\Theta )$, called the symmetric logarithmic derivative (SLD), by $\frac{\partial\rho_{\Theta}}{\partial\theta^{i}}=\frac{1}{2}({\rm L}_{i}^{S}(\Theta )\,\rho_{\Theta }+\rho_{\Theta}\,{\rm L}_{i}^{S}(\Theta))$. SLD ${\rm L}_{i}^{S}(\Theta)$ can be considered as the quantum analogue of the logarithmic derivative. Then the SLD Fisher information is given by $J_{ij}^{S}(\Theta)=\frac{1}{2}{\rm Tr}[\rho_{\Theta}({\rm L}_{i}^{S}(\Theta){\rm L}_{j}^{S}(\Theta)+{\rm L}_{j}^{S}(\Theta){\rm L}_{i}^{S}(\Theta))]$.  See, for example, Refs.~\cite{Holevo2001,ANtext2000}.

\bibitem{note_Fisher}
The Fisher information matrix, ${\rm J}(\Theta) \equiv [J_{ij}(\Theta)]$, is defined by $J_{ij}(\Theta)=\sum_{n_{1}=0}^{\infty}\!\cdots\!\sum_{n_{16}=0}^{\infty}P_{0}(N){\rm L}^{C}_{i}(\Theta){\rm L}^{C}_{j}(\Theta)$, where $P_{0}(N) \equiv P(N|\Theta_{0})$ and $L^{C}_{i}(\Theta) \equiv \frac{\partial}{\partial\theta^{i}}\ln[P(N|\Theta)]$. Here, $P(N|\Theta)=\prod_{\nu=1}^{16}e^{-M_{\nu}} \frac{M_{\nu}^{n_{\nu}}}{n_{\nu}!}$ is the probability density function obtaining a certain set of coincidence counts $\{N\}$, where $M_{\nu}=C tTr[|m_{\nu} \rangle\langle m_{\nu}|\rho_{\Theta}]$. $C$ is called the nuisance parameter; it corresponds to the coincidence counting rate without any polarizers ($|m_{\nu}\rangle \langle m_{\nu}|=I$). See, for example, Refs.~\cite{CTtext1991,ANtext2000}.

\bibitem{Braunstein1992}  
S.~L.~Braunstein, J.~Phys.~A:~Math.~Gen.~%
{\bf 25}, 3813 (1992).

\bibitem{note_findminimum}
This task was executed by the {\it FindMinimum} function of MATHEMATICA 4.0 (employing the multi-dimensional Powell algorithm), as in Ref.~\cite{JKMW2001}. Note that the minimum found by this function is not necessarily the {\it global} minimum.

\bibitem{note_parameters}
Although the number of independent parameters for the rank-4, 3, 2, and 1 density matrices of two-qubit states are 15, 14, 11, and 6, respectively, {\it a priori} unknown parameter $C$ is also included here. 

\bibitem{GM2000}  
R.~D.~Gill and S.~Massar, Phys.~Rev.~A~{\bf 61}
042312 (2000).


\end{thebibliography}
\end{document}